\documentclass[twocolumn,preprintnumbers,amsmath,aps]{revtex4}
\usepackage{booktabs}
\usepackage{graphicx}
\usepackage{epstopdf}
\usepackage{dcolumn}
\usepackage{bm}
\usepackage{color}
\usepackage{multirow}
\begin{document}
\def\eq#1{(\ref{#1})}
\def\fig#1{\hspace{1mm}Fig. \ref{#1}}
\def\tab#1{\hspace{1mm}\ref{#1}}
{\color{black}
\title{Scalability of non-adiabatic effects in lithium-decorated graphene superconductor}

\author{Dominik Szcz{\c{e}}{\'s}niak}

\affiliation{Department of Theoretical Physics, Faculty of Science and Technology, Jan D{\l}ugosz University in Cz{\c{e}}stochowa, 13/15 Armii Krajowej Ave., 42200 Cz{\c{e}}stochowa, Poland}
\date{\today} 
\begin{abstract}

The analysis is conducted to unveil how the non-adiabatic effects scale within the superconducting phase of lithium-decorated graphene (LiC$_{6}$). Based on the Eliashberg formalism it is shown that the non-adiabatic effects notably reduce essential superconducting parameters in LiC$_{6}$ and arise as a significant oppressor of the discussed phase. Moreover, nonadiabaticity is found to scale with the strength of superconductivity, proportionally to the phonon energy scale and inversely with respect to the electron-phonon coupling. These findings are partially in contrast to other theoretical studies and show that superconductivity in LiC${_6}$ is more peculiar than previously anticipated. In this context, the guidelines for enhancing superconducting phase in LiC${_6}$ and sibling materials are also proposed.

\end{abstract}

\maketitle

\section{Introduction}

In the literature, a number of scenarios exist on how to potentially induce conventional superconductivity in graphene, promising novel applications of this intriguing carbon allotrope \cite{lu, thingstad, durajski, uchihashi, zheng, zhou, pesic, guzman, kaloni, profeta, einenkel, savini}. These strategies are mainly aimed at the structural modifications of graphene to enhance number of charge carriers at the Fermi level and alter intrinsic semimetallic character of this material. Among the resulting structures, lithium-decorated graphene (LiC$_{6}$) is here of particular interest since it is suggested to host phonon-mediated superconducting state generated via process analogous to the intercalation of graphite \cite{profeta}. This approach relies on the introduction of adatoms that break chiral symmetry of graphene and lifts Fermi level to the van Hove singularity \cite{gholami, bao, gutierrez, profeta}. However, superconductivity in LiC$_{6}$ not only derives from the well-established method of synthesis but also appears to be actually feasible, as partially confirmed within the experiment \cite{ludbrook}.

According to the above, LiC$_{6}$ may be considered as a proving ground for the phonon-mediated superconductivity at low-dimensions. Indeed, in recent years the superconducting phase in LiC$_{6}$ received notable attention in terms of its fundamental properties. The related studies were devoted, but not limited to, the role of strong electron-phonon coupling \cite{szczesniak1}, character of superconducting gap \cite{zheng}, symmetry-breaking effects \cite{gholami}, ways of enhancing superconducting state \cite{pesic}, or the substrate ievmpact on superconductivity \cite{kaloni}. Beside listed directions of research, LiC$_{6}$ appears also to be a perfect example of superconducting material for studying the influence of non-adiabatic pairing \cite{szczesniak2}. This aspect is particularly intriguing since non-adiabatic effects tend to manifest themselves relatively rarely in conventional superconductors. The reason for that relates to the non-comparable electronic and phononic energy scales in most superconducting materials with the electron-phonon pairing mechanism \cite{pietronero, grimaldi1, grimaldi2}. In other words, it can be often assumed that electrons follow adiabatically ionic oscillations and that the corresponding superconducting state can be described in a self-consistent manner \cite{hu, capelluti1}. However, this is not the case when superconducting materials exhibit shallow conduction band such as the fullerenes \cite{paci}, fullerides \cite{capelluti1}, bismuthates \cite{szczesniak3}, transition-metal-oxides \cite{gorkov} or the discussed LiC$_{6}$ \cite{szczesniak2} (see \cite{talantsev} for the review of nonadiabatic superconductors).

So far, the characteristic energy scales are one of the few signatures of non-adiabatic superconductivity in LiC$_{6}$. Other than that recent theoretical studies show non-adiabatic effects to notably reduce magnitude of depairing interaction in the discussed material, in comparison to the adiabatic regime \cite{szczesniak2}. These findings are also supplemented by the considerations suggesting that non-adiabaticity contributes to the electron-phonon coupling ($\lambda$) and modulates the transition temperature ($T_{C}$) in doped graphene \cite{hu} or two-dimensional superconductors in general \cite{schrodi}. Still, little is known about the scalability of non-adiabatic effects in LiC$_{6}$. In the first approximation, it can be only qualitatively argued that their impact changes according to the Migdal's ratio (known also as the expansion ratio) given by $m=\lambda \omega_{D}/E_{F}$, where $E_{F}$ is the Fermi energy and $\omega_{D}$ denotes Debye's frequency \cite{pietronero, grimaldi1, grimaldi2}. In fact, although $m$-ratio can provide some information on the scalability problem it is mostly used to determine whether or not given material can be described within the Migdal's theorem \cite{migdal} {\it i.e.} within adiabatic or non-adiabatic regime. Hence, the measurable and direct role of the above parameters and their variations in shaping non-adiabatic superconductivity in LiC$_{6}$ is somewhat hindered. In details, it is unknown how changes in the $m$-ratio components modify experimentally observable thermodynamic properties such as the superconducting gap or the transition temperature. This is to say, what trends in thermodynamics can be expected due to the strength of the non-adiabatic effects. As a result, the relevancy of the energy scales and the electron-phonon coupling in the non-adiabatic limit is also not well-estimated yet. Therefore, addressing these aspects would be of great importance to the better understanding of superconductivity in LiC$_{6}$ and potentially other sibling low-dimensional materials. Moreover, it should also help in assessing impact of the external factors that can be applied to modify the aforementioned properties and ultimately enhance the superconducting state even further.

To provide deeper insight into the scalability of non-adiabatic effects in LiC$_{6}$, the present study analyzes behavior of the discussed material in the adiabatic and non-adiabatic limit when the expansion ratio parameters vary. This is done within the Eliashberg formalism that generalizes conventional Bardeen-Cooper-Schrieffer (BCS) theory of superconductivity \cite{bardeen1, bardeen2} by incorporating the strong-coupling, retardation and non-adiabatic effects \cite{eliashberg, carbotte, pietronero, grimaldi1, grimaldi2}. As a result it allows to consider both regimes of interest and relate predictions on the pivotal thermodynamics to the potential factors responsible for the $m$-ratio variations. Here the latter is modeled in reference to \cite{pesic}, by recalling the fact that the deformation potential is able to simultaneously influence energy scales and the electron-phonon coupling in a given superconducting material. This effect is captured via the percentage change in graphene lattice constant given as $\delta = \left|a - a_{0} \right| / a_{0} \times 100 \%$, where $a_{0}(a)$ is the unmodified (modified) lattice constant value. In what follows, several levels of $\delta$ are considered allowing for tracing changes in the $m$-ratio components on the same footing and in the direct relation to the experimentally observable case ($\delta=0 \%$). Based on that it is possible to unveiled how the non-adiabatic effects scale in LiC$_{6}$ and what can be done to eliminate their potentially negative consequences.

\section{Metodology}

The theoretical formalism of choice is provided here by following the study of Freericks {\it et al.} \cite{freericks}, where convenient form of the generalized Eliashberg equations for considering the non-adiabatic superconductivity is presented. This theoretical approach is based on the perturbative theory introduced originally by Pietronero {\it et al.} in \cite{pietronero, grimaldi1, grimaldi2}, which incorporates non-adiabatic effects via vertex corrections to the electron-phonon interaction. However, the theoretical scenario given in \cite{freericks} includes specific computational techniques for better accuracy and efficiency, such as the perturbative theory on the imaginary-axis and the high-frequency resummation schemes. In this manner, the resulting equations provide compromise between predictive capabilities and computational requirements. Note that such approach was already proved successful in describing non-adiabatic superconductivity not only in LiC$_{6}$ but also other phonon-mediated superconductors such as lead \cite{freericks} or bismuthates \cite{szczesniak4}.

In respect to the above, inital approximations are assumed in accordance to \cite{freericks} and the character of the superconducting phase in LiC$_{6}$. In particular, (i) the direct dependence on momentum is neglected for the electron-phonon matrix elements, in correspondence to the isotropic nature of superconducting gap in LiC$_{6}$, (ii) the depairing correlations are modeled only by the first-order Coulomb pseudopotential terms, due to the fact that higher-order contributions are negligibly small for phonon-mediated superconductors, (iii) similarly only the lowest-order vertex corrections to the electron-phonon interaction are considered to describe the non-adiabatic effects, since the Fermi liquid picture in LiC$_{6}$ appears to be conserved. As a results, it is possible to derive self-consistent Eliashberg equations beyond Midal's theorem within perturbation scheme. In details, their form on the imaginary axis for the order parameter function ($\phi_{n}=\phi\left(i\omega_{n}\right)$) and the wave function renormalization factor ($Z_{n}= Z\left(i\omega_{n}\right)$) is following:
\begin{equation}
\label{eq01}
\phi_{n}=\pi k_{B}T\sum_{m=-M}^{M}
\frac{K_{n,m}-\mu_{m}^{\star}}
{\sqrt{\omega_m^2Z^{2}_{m}+\phi^{2}_{m}}}\phi_{m} - V_{\phi},
\end{equation}
\begin{equation}
\label{eq02}
Z_{n}=1+\frac{\pi k_{B}T}{\omega_{n}}\sum_{m=-M}^{M}
\frac{K_{n,m}}{\sqrt{\omega_m^2Z^{2}_{m}+\phi^{2}_{m}}}\omega_{m}Z_{m} - V_{Z},
\end{equation}
where, $k_{B}T$ is the inverse temperature, with $k_{B}$ denoting the Boltzmann constant. In what follows, $\omega_{n}=\pi k_{B}T\left(2n+1\right)$ is the $n$-th Matsubara frequency with the cutoff $M=1100$ for numerical stability above $T=2$ K. Moreover, $K_{n,m}$ stands for the electron-phonon pairing kernel given as:
\begin{equation}
\label{eq03}
K_{n,m}=2\int_0^{\omega_{D}}d\omega\frac{\omega}{\omega ^2+4\pi^{2}\left(k_{B}T\right)^{2}\left(n-m\right)^{2}}\alpha^{2}F\left(\omega\right),
\end{equation}
with $\omega$ being the phonon frequency, $\alpha$ describing the average electron-phonon coupling and $F\left(\omega\right)$ denoting the phonon density of states. Note that the product of the two latter is known as the electron-phonon spectral function which provides most important information about a physical system within the Eliashberg formalism \cite{carbotte}. Here, several $\alpha^{2}F\left(\omega\right)$ functions are considered, each of them corresponding to the different $\delta$-value in order to analyze behavior of LiC$_6$ when the Migdal's parameter and its components very. For this purpose, the exact forms of the $\alpha^{2}F\left(\omega\right)$ functions are assumed after \cite{pesic} for $\delta \in \left<0, 3, 5, 7, 10 \right>\%$, where the first case corresponds to the experimentally observed superconducting phase of LiC$_6$ whereas the remaining functions describe potential variations from its pristine form. The remaining information is given via the Coulomb pseudopotential $\mu_{n}^{\star}=\mu^{\star}\theta \left(\omega_{c}-|\omega_{n}|\right)$, with $\theta$ standing for the the Heaviside function and $\omega_{c}$ for the cut-off frequency. To consider all the $\delta$ cases on equal footing the conventional value of $\mu^{\star}=0.1$ \cite{bauer} which is close to the magnitude of Coulomb depairing interaction predicted for LiC$_6$ at $\delta=0\%$ \cite{szczesniak2}.

Finally, $V_{\phi}$ and $V_{Z}$ are the lowest-order vertex correction terms of the following form:
\begin{eqnarray}
\label{eq04}
\nonumber
V_{\phi}&=&\frac{\pi^{3}\left(k_{B}T\right)^{2}}{4E_{F}} \sum_{m=-M}^{M}\sum_{m'=-M}^{M} K_{n,m}K_{n,m'}
\\ \nonumber
&\times&\frac{1}
{\sqrt{\left(\omega_m^2Z^{2}_{m}+\phi^{2}_{m}\right)\left(\omega_{m'}^2Z^{2}_{m'}+\phi^{2}_{m'}\right)}}
\\ \nonumber
&\times&\frac{1}
{\sqrt{\left(\omega_{m''}^2Z^{2}_{m''}+\phi^{2}_{m''}\right)}}
\\ \nonumber
&\times&\left(\phi_{m}\phi_{m'}\phi_{m''}+2\phi_{m}\omega_{m'}Z_{m'}\omega_{m''}Z_{m''} \right.
\\
&-&\left.\omega_{m}Z_{m}\omega_{m'}Z_{m'}
\phi_{m''}
\right),
\end{eqnarray}
and
\begin{eqnarray}
\label{eq05}
\nonumber
V_{Z}&=& \frac{\pi^{3}\left(k_{B}T\right)^{2}}{4E_{F}\omega_{n}} \sum_{m=-M}^{M}\sum_{m'=-M}^{M} K_{n,m}K_{n,m'}
\\ \nonumber
&\times&
\frac{1}
{\sqrt{\left(\omega_m^2Z^{2}_{m}+\phi^{2}_{m}\right)
       \left(\omega_{m'}^2Z^{2}_{m'}+\phi^{2}_{m'}\right)}}
\\ \nonumber
&\times&\frac{1}
{\sqrt{\left(\omega_{m''}^2Z^{2}_{m''}+\phi^{2}_{m''}\right)}}
\\ \nonumber
&\times&\left(\omega_{m}Z_{m}\omega_{m'}Z_{m'}\omega_{m''}Z_{m''}+2\omega_{m}Z_{m}\phi_{m'}\phi_{m''} \right.
\\
&-&\left.\phi_{m}\phi_{m'}\omega_{m''}Z_{m''}\right).
\end{eqnarray}
Based on the above, when vertex corrections are considered within the Eqs. (\ref{eq01}) and (\ref{eq02}) they are refereed here to as the non-adiabatic Eliashberg equations (N-E), otherwise, when the corrections are neglected, the formalism is reduced to the adiabatic Eliashberg equations (A-E).

In what follows, by solving Eqs. (\ref{eq01}) and (\ref{eq02}) it is possible to obtain estimates on the most important thermodynamic properties of superconducting state in LiC$_{6}$. Specifically, the central role in such analysis is played by the order parameter function that is obtained from Eqs. (\ref{eq01}) and (\ref{eq02}) as: $\Delta_{n}(T)=\phi_{n}/Z_{n}$. Here of special interest is the maximum value ($m=1$) of $\Delta_{n}(T)$ which contains information on the transition temperature and the superconducting gap half-width. The former is determined based on the relation $\Delta_{m=1}(T_{C})=0$, whereas the latter is given by $\Delta_{m=1}(T_{0})$, with $T_{0}=2$ K being the lowest temperature assumed for calculations. Since the aforementioned $\alpha^{2}F\left(\omega\right)$ function is dependent on the characteristic energy scales and the electron-phonon coupling constant, the described solutions of the Eliashberg equations also inherit such dependence, allowing for the analysis of interest.

\section{The results and discussion}

%
\begin{figure}[ht!]
\includegraphics[width=\columnwidth]{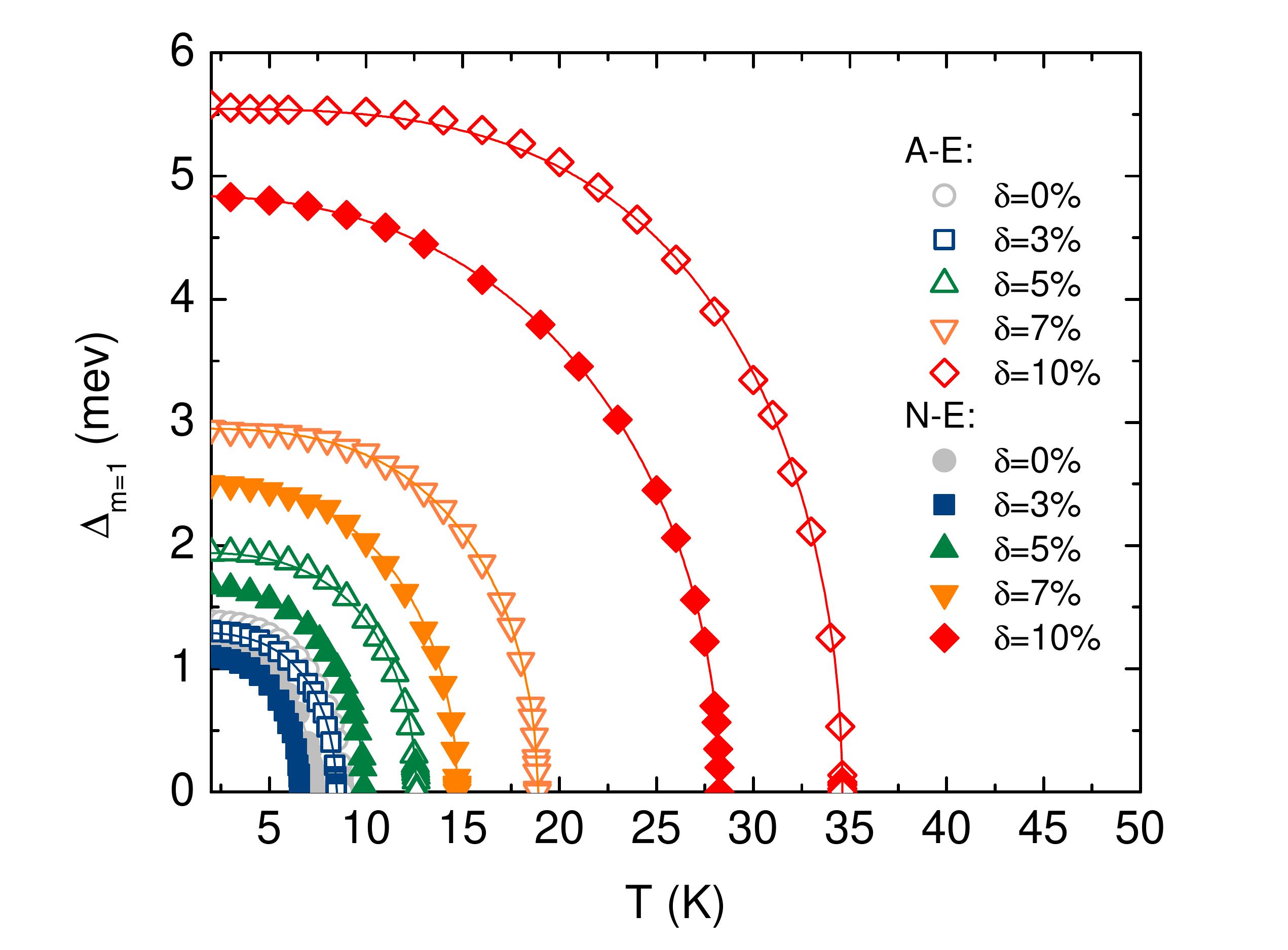}
\caption{The maximum value of order parameter ($\Delta_{m=1}(T)$) as a function of temperature in LiC$_{6}$. The results are depicted for the selected values of lattice constant deviation in graphene ($\delta$) as obtained within the adiabatic (closed symbols) and non-adiabatic (open symbols) regime of the Eliashberg equations. Solid lines constitute the guides for an eye.}
\label{fig01}
\end{figure}

In Fig. \ref{fig01}, the main numerical results are presented as obtained by solving Eqs. (\ref{eq01}) and (\ref{eq02}) iteratively with respect to the temperature (see \cite{szczesniak3} and \cite{freericks} for more details on the computational methods used here). In details, Fig. \ref{fig01} depicts the behavior of $\Delta_{m=1}(T)$ function for $T\in\left<T_{0},T_{C}\right>$ at the assumed levels of lattice constant deviation denoted by $\delta$. Note that the 0\% case corresponds to the unaltered LiC$_{6}$ material, hosting the experimentally observable superconducting phase. On the other hand, the remaining cases describe situation when crystal lattice of graphene changes according to the already mentioned expression: $\delta = \left|a - a_{0} \right| / a_{0} \times 100\%$, with $a_{0}(a)$ standing for the unmodified (modified) lattice constant. Moreover, as allowed by the employed formalism, the discussed thermal behavior of $\Delta_{m=1}(T)$ function is plotted for the adiabatic (open symbols) and non-adiabatic (closed symbols) regime. In all figures, symbols relate to the exact numerical results of the Eliashberg equations and solid lines constitute guides for an eye.

The result depicted in Fig. \ref{fig01} reveal several general aspects of the superconducting state in LiC$_{6}$. In particular, it can be observed that for $\delta>3\%$ the increase of the $\delta$ value causes notable increase of the $\Delta_{m=1}(T)$ in the entire temperature range for both considered regimes. This trend is not conserved only when comparing result obtained at the two lowest levels of $\delta$, as caused by the $\delta$-driven increase of charge transfer that empties interlayer states and notably reduces the electron-phonon coupling constant at $\delta=3\%$ with respect to the 0\% case \cite{pesic}. Nonetheless, the observed effect means that above some level of $\delta$ the superconducting state in LiC$_{6}$ is clearly enhanced. This observation can be quantified by deducing the transition temperature values from the obtained results. In particular, $T_{C}=8.78$ K at $\delta=0\%$ and $T_{C}\in\left<8.48, 34.61 \right>$ K for $\delta\in \left<3,10\right>\%$ within the A-E limit, whereas $T_{C}=7.29$ K at $\delta=0\%$ and $T_{C}\in\left<6.59, 28.46 \right>$ K for $\delta\in \left<3,10\right>\%$ when considering the N-E equations. Note that these observations are in qualitative agreement with the previous studies conducted within the adiabatic limit by using the Allen-Dynes formula in \cite{pesic}. The difference between these data sets and results given in \cite{pesic} is due to the fact that the assumed Eliashberg equations incorporate strong-coupling, retardation, and non-adiabatic effects which are missing in the Allen-Dynes formula. At this point, it is also instructive to note that the $T_{C}$ value estimated at $\delta=0\%$ is slightly higher in comparison to the predictions made within the Eliashberg formalism for the experimentally derived electron-phonon spectral function, as presented in \cite{szczesniak2}. This discrepancy is obviously caused by the assumed value of $\mu^{\star}$, smaller than the one suggested in \cite{szczesniak2}. The reason to make such assumption is to allow for better comparison not only with the BCS-derived results given in \cite{pesic} but also other two-dimensional superconductors, which are often still hypothetical structures and their superconducting state is described by $\mu^{\star} \sim 0.1$ (see {\it e.g.} \cite{wan, ge}). Note that even if $\mu^{\star}$ would be assumed here after \cite{szczesniak2}, the main outcomes and findings of the present analysis would not change. This includes estimates on the superconducting gap half-width that can be made based on the results plotted in Fig. \ref{fig01}. This is to say, the general behavior of $\Delta_{m=1}(T_{0})$ parameter is the same as in the case of $T_{C}$ and it will not change qualitatively when assuming other $\mu^{\star}$ value. Specifically, in the present study $\Delta_{m=1}(T_{0})=1.39$ meV at $\delta=0\%$ whereas $\Delta_{m=1}(T_{0})\in\left<1.30, 5.55 \right>$ meV for $\delta\in \left<3,10\right>\%$ in the A-E regime, while the N-E equations yield $\Delta_{m=1}(T_{0})=1.22$ meV at $\delta=0\%$ and $\Delta_{m=1}(T_{0})\in\left<1.11, 4.84 \right>$ meV for $\delta\in \left<3,10\right>\%$. Note that results on $T_{C}$ and $\Delta_{m=1}(T_{0})$ can be supplemented by introducing their characteristic ratio, familiar in the BCS theory and given by \cite{bardeen1, bardeen2, carbotte}:
\begin{equation}
\label{eq06}
R=\frac{2\Delta_{m=1}(T_{0})}{k_{B}T_{C}}.
\end{equation}
The Eq. (\ref{eq06}) not only allows for additional insight into the considered problem but also provides yet another observable for future comparisons with the experiment. As it can be expected, the obtained values of $R$ follow the same trends like the $T_{C}$ and $\Delta_{m=1}(T_{0})$ parameters. The values of $R$ base on the A-E equations are $R=3.67$ at $\delta=0\%$ and $R\in\left<3.55, 3.72 \right>$ for $\delta\in \left<3,10\right>\%$. On the other hand, the N-E equations give $R=3.89$ at $\delta=0\%$ and $R\in\left<3.92, 3.98 \right>$ for $\delta\in \left<3,10\right>\%$. Still, both sets present values higher than the level suggested within the BCS theory and equal to 3.53. It means that the strong-coupling and retardation effects play relatively important role in shaping the superconducting state in LiC$_{6}$. This observation is in agreement with the strength of the electron-phonon coupling reported in \cite{pesic} and previous findings given in \cite{szczesniak2}.

\begin{figure}[ht!]
\includegraphics[width=\columnwidth]{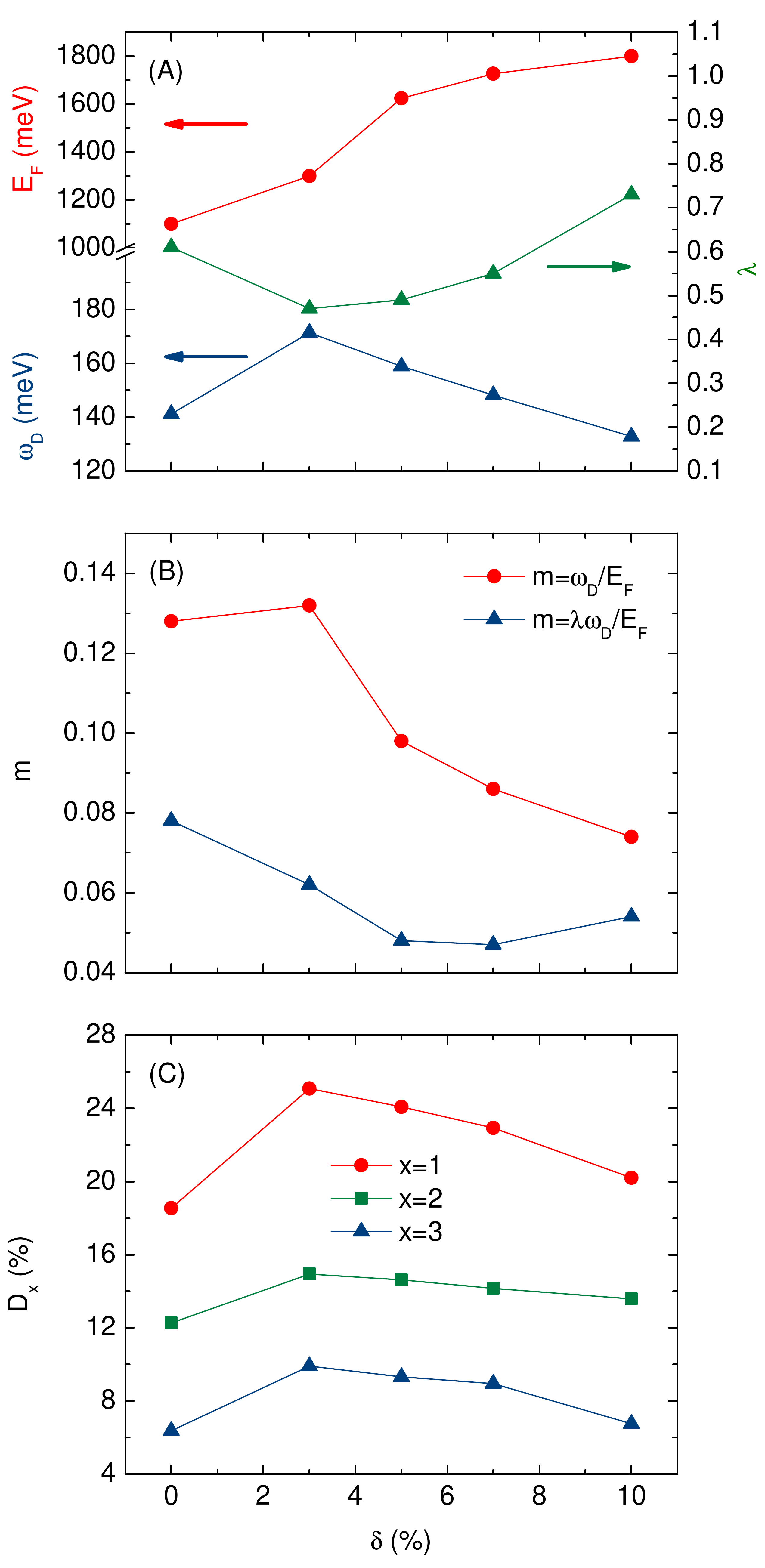}
\caption{The behavior of (A) the Fermi energy ($E_{F}$), Debye's frequency ($\omega_{D}$) and electron-phonon coupling constant ($\lambda$), (B) the dressed  ($m=\lambda \omega_{D}/E_{F}$) and bare ($m=\omega_{D}/E_{F}$) Migdal's ratio as well as (C) the percentage differences between adiabatic and non-adiabatic estimates for the critical temperature ($D_{1}$), superconducting gap half-width ($D_{2}$) and their cumulative ratio ($D_{3}$) at the selected values of the lattice deviation in graphene ($\delta$) in the LiC$_{6}$ superconductor. The closed symbols depicts exact results, the solid lines are the guides for an eye and the color arrows points to the corresponding axes.}
\label{fig02}
\end{figure}
\begin{table*}
\caption{The parameters of superconducting state in LiC$_{6}$ for the selected values of the lattice deviation in graphene ($\delta$). In a consecutive order, the parameters are: the electron-phonon coupling constant ($\lambda$), the Debye's frequency ($\omega_{d}$), the Fermi energy ($E_{F}$), the bare ($\omega_{d}/E_{F}$) and dressed ($\lambda \omega_{d}/E_{F}$) Migdal's ratio, the transition temperature ($T_{C}$), the superconducting gap half-width ($\Delta_{m=1}(T_{0})$) as well as the thermodynamic ratio for the two last ones ($R$). Note that, where necessary, the results are presented for the in terms of the adiabatic ($A-E$) and non-adiabatic ($N-E$) regime. Moreover, the percentage differences between estimates in these two limits for $T_{C}$ ($D_{1}$), $\Delta_{m=1}(T_{0})$ ($D_{2}$) and $R$ ($D_{3}$) are also given.}
\label{tab01}
\begin{ruledtabular}
\begin{tabular}{c | c c c c c | c c c | c c c | c c c}
& & & & & & & A-E & & & N-E & \\
\hline
$\delta$ & $\lambda$ & $\omega_{d}$ & $E_{F}$ & $\omega_{d}/E_{F}$ & $\lambda \omega_{d}/E_{F}$ & $T_{C}$ & $\Delta_{m=1}(T_{0})$ & R & $T_{C}$ & $\Delta_{m=1}(T_{0})$ & R & $D_{1}$ & $D_{2}$ & $D_{3}$ \\
(\%) &  & (meV) & (meV) &  &  & (K) & (meV) &  & (K) & (meV) &  & (\%) & (\%) & (\%) \\
\hline
0 & 0.61 & 141.21 & 1100 & 0.128 & 0.078 & 8.78 & 1.39 & 3.67 & 7.29 & 1.22 & 3.89 & 18.54 & 13.03 & 6.37\\
3 & 0.47 & 171.34 & 1300 & 0.132 & 0.062 & 8.48 & 1.30 & 3.55 & 6.59 & 1.11 & 3.92 & 25.08 & 15.77 & 9.91 \\
5 & 0.49 & 158.82 & 1624 & 0.098 & 0.048 & 12.61 & 1.94 & 3.58 & 9.90 & 1.68 & 3.93 & 24.08 & 14.36 & 9.32 \\
7 & 0.55 & 148.16 & 1726 & 0.086 & 0.047 & 18.86 & 2.95 & 3.63 & 14.98 & 2.56 & 3.97 & 22.93 & 14.16 & 8.95 \\
10 & 0.73 & 132.79 & 1800 & 0.074 & 0.054 & 34.61 & 5.55 & 3.72 & 28.26 & 4.84 & 3.98 & 20.20 & 13.67 & 6.75
\end{tabular}
\end{ruledtabular}
\end{table*}

Beside the above observations, it is also crucial to note that the reported results suggest simultaneous changes in the energy scales and the electron-phonon coupling constant due to the variations of $\delta$ parameter. Indeed, all of these characteristic parameters exhibit increasing or decreasing trends along with the growing $\delta$ value (see Fig. \ref{fig02} (A)). For convenience, their cumulative behavior is depicted in Fig. \ref{fig02} (B) in terms of already introduced dressed Migdal's ratio ($m=\lambda \omega_{D}/E_{F}$) but also its bare counterpart ($m=\omega_{D}/E_{F}$). In what follows, it is argued here that the scalability of non-adiabatic effects in LiC$_{6}$ can be traced with respect to the pivotal parameters entering Migdal's ratio. In this context, first it should be noted that for each considered $\delta$ value the non-adiabatic equations yield lower $\Delta_{m=1}(T)$ values than their adiabatic counterparts (see Fig. \ref{fig01}). This directly relates to the fact that the transition temperature values as well as the estimates of the superconducting gap are lower in the non-adiabatic regime when comparing to the adiabatic one. As a results, the percentage difference between estimates made in two considered regimes can be introduced as a measure of non-adiabatic effects impact on superconducting phase in LiC$_{6}$. This new measure is depicted for all considered thermodynamic parameters in Fig. \ref{fig02} (C). In details, the percentage difference between $T_{C}$ values determined in the adiabatic and non-adiabatic limits is $D_{1}=18.54\%$ at $\delta=0$\% and $D_{1}=\left< 25.08, 20.20 \right>$\% for $\delta\in \left<3,10\right>\%$. Similarly, the same percentage measure but for the $\Delta_{m=1}(T_{0})$ parameter is $D_{2}=13.03\%$ for $\delta=0$\% and $D_{2}=\left< 15.77, 13.67 \right>$\% when $\delta\in \left<3,10\right>\%$. Finally, the cumulative ratio gives the corresponding percentage $D_{3}=6.37\%$ for $\delta=0$\% and $D_{3}=\left< 9.91, 6.75 \right>$\% when again $\delta\in \left<3,10\right>\%$. Based on these results, it is clear that the critical temperature is the most influenced by the non-adiabatic effects from all three considered parameters. It also presents the most visible signature of the charge transfer from interlayer states at $\delta\in 3\%$. On the contrary, the cumulative ratio shows the smallest discrepancies. Still all three parameters exhibit the same qualitative behavior {\it i.e.} the percentage difference between adiabatic and non-adiabatic results increases up to $\delta\in 3\%$ and then starts to almost linearly decrease as the $\delta$ takes higher values.

The final observations can be made when comparing results presented in Fig. \ref{fig02} (C) with the estimates depicted in Figs. \ref{fig02} (A) and (B). In particular, it can qualitatively argued that the behavior of percentage measures does not fully resemble the Migdal's ratio dependence on the $\delta$ value, although the latter is considered to be the first approximation approach to provide information on the scalability of non-adiabatic effects in a superconductors. Precisely speaking, only the bare ratio can be considered somewhat similar in behavior to the percentage measures, whereas its dressed value presents almost inverse character with respect to the parameters given in Fig. \ref{fig02} (C). To inspect these discrepancies even further it is instructive to compare the percentage difference measures with the characteristic component parameters of the Migdal's ratio, as plotted in Figs. \ref{fig02} (A). The outcome is that only the Debye's energy scales qualitatively the same as the percentage difference measures, while the electron-phonon coupling gives inverse characteristic and the electronic energy scale is practically nowhere similar to the results given in Figs. \ref{fig02} (C).

\section{Summary and conclusions}

In summary, the presented analysis provides new insight into the superconducting properties of LiC${_6}$ in terms of its non-adiabatic characteristic. It shows how the non-adiabatic effects scale in LiC${_6}$ with respect to the deviation of lattice constant in graphene, which can be considered as an exemplary factor that modifies strength of the superconducting state. In details, the discussed scalability is expressed here in terms of the percentage difference between estimates of pivotal thermodynamic parameters (the transition temperature, superconducting gap and their ratio) obtained within the adiabatic and non-adiabatic regime, allowing for further comparison with the Migdal's expansion ratio that characterizes nonadiabaticity in the first approximation. For convenience, all the obtained numerical results are summarized in Tab. \ref{tab01}.

Based on the above findings it is possible to draw several conclusions related, but no limited to, the scalability of non-adiabatic effects in LiC$_{6}$. In details:

\begin{itemize}

\item[(i)]{The introduction of vertex corrections to the electron-phonon interaction causes notable changes in the pivotal thermodynamic parameters of the superconducting state in LiC{$_6$}, in particular their decrease in comparison to the adiabatic limit (see Fig. \ref{fig02} (C)). This is to say, the superconducting state appears to have strongly non-adiabatic character, where non-adiabatic effects act as an important oppressor of superconductivity in LiC{$_6$}. Notably the superconducting state sustains its non-adiabatic character even when superconductivity in LiC$_{6}$ is strongly enhanced. Still the non-adiabatic effects are observed to visibly vary with the strength of superconducting phase. Moreover, it is found that nonadiabaticity is supplemented by the strong-coupling and retardation effects, meaning that LiC$_{6}$ is a somewhat unorthodox phonon-mediated superconductor.}

\item[(ii)]{The considered thermodynamic parameters present the same qualitative behavior under the influence of non-adiabatic effects when the strength of superconductivity is varied (see Fig. \ref{fig02} (C)). Nonetheless, the critical temperature is suggested to be particularly sensitive to nonadiabaticity, while superconducting gap is showing much smaller dependence on the variation of the discussed effects. As a result, this opens new prospect for increasing transition temperature value in LiC${_6}$ according to the presented here findings, saying that smaller magnitude of non-adiabatic effects leads to the higher transition temperature (see Tab. \ref{tab01}). Note that this trend is not conserved when including results for the unaltered LiC$_{6}$, due to the decreased charge transfer from the interlayer states in comparison to other considered cases. It can be additionally argued that such trend may be considered general for other graphene-based superconductor which exhibit similar dependence on nonadiabaticity (see {\it e.g.} recent study on the electron-doped graphene \cite{szczesniak5}).}

\item[(iii)]{The deeper inspection of the obtained results shows that the non-adiabatic effects scale with the strength of superconductivity, proportionally to the phonon energy scale and inversely to the electron-phonon coupling magnitude (see Figs. \ref{fig02} (A) and (C)). Note that, while the former observation agrees with the predictions of both considered forms of the Migdal's ratio, the latter is in contrast to what can be expected based on the dressed parameter and previous theoretical studies considering superconductivity in two-dimensional systems \cite{hu, schrodi}. However, the mentioned trends does not take into account existing interplay between all components of the Migdal's ratio and the fact that the electron-phonon coupling constant increases almost linearly with the electronic energy scale (see Tab. \ref{tab01}). As a results, strong electron-phonon correlations correspond to the relatively wide conduction band that causes suppression of nonadiabaticity. This argument is confirmed qualitatively by the observed here cumulative behavior of the Migdal's ratio (see Figs. \ref{fig02} (B)). This is to say the electron-phonon coupling cannot be always considered to be improved in graphene-based superconductors by the non-adiabatic effects as previously suggested in \cite{hu, schrodi}.}

\end{itemize}

To sum up, the perspectives for future research can be given. In details, to provide better understating of the superconducting state in LiC$_{6}$ the discussed non-adiabatic effects should be considered beyond the isotropic approximation. Note that such preliminary analysis can be already found in \cite{schrodi}, while the adiabatic anisotropic investigations are available in \cite{zheng}. The present study can be also extended further toward other experimentally observable thermodynamic parameters such as the free energy or the critical thermodynamic field, according to their importance in discussing the non-adiabatic effects \cite{miller}. Finally, recent discussion given in \cite{talantsev} suggest strongly metallic behavior of LiC${_6}$ despite its high value of the Migdal's ratio. In other words, the superconducting state in LiC${_6}$ may appear to be more peculiar than previously anticipated and additional investigations in this directions should be of great interest.

\bibliography{bibliography}
\end{document}